# Zitterbewegung and its significance for Hawking radiation

**Zhi-Yong Wang[*], Cai-Dong Xiong, Qi Qiu**

*E-mail:  zywang@uestc.edu.cn

*School of Optoelectronic Information, University of Electronic Science and Technology of China,*

*Chengdu 610054, CHINA*

**Abstract**

An old interest in the zitterbewegung (ZB) of the Dirac electron has recently been rekindled by the investigations on spintronics and graphene, etc. In this letter, we study the ZB of the Dirac electron at the level of quantum field theory, and reveal another interesting aspect of the ZB: one can depict Hawking radiation in terms of ZB.

**1. Introduction**

Recently, an old interest in the zitterbewegung (ZB) of the Dirac electron has been rekindled by the investigations on spintronics, graphene, and superconducting systems, etc. [1-9]. In particular, there are important progresses in improving the predictions for detecting ZB and relating them to Schrodinger cats in trapped ions [10, 11]. On the other hand, as is well known, the Hawking radiation can be viewed from very different perspectives. In this letter, we present another perspective on the Hawking radiation by means of ZB. In the following the natural units of $\hbar = c = 1$ are applied, and the four-dimensional (4D) flat space-time metric tensor is taken as $\eta_{\mu\nu} = \text{diag}(1,-1,-1,-1)$ ($\mu,\nu = 0,1,2,3$). Repeated indices must be summed according to the Einstein rule.

## 2. Quantized-field description of the Dirac electron's ZB

Traditionally, the investigations on ZB have been developed at the level of quantum mechanics, where one has to resort to the concept of position operator. However, at the level of quantum field theory, the ZB should be studied in terms of a 3D current vector (or 3D momentum vector). The free Dirac field can be expressed as

$$\psi(x) = \frac{1}{\sqrt{V}} \sum_{p,s} [\hat{c}(p,s)u(p,s)\exp(-ip_\mu x^\mu) + \hat{d}^\dagger(p,s)v(p,s)\exp(ip_\mu x^\mu)], \qquad (1)$$

where $V = \int d^3x$ represents the normalization volume (the box normalization is adopted in this letter), $s = 1,2$ correspond to the spins of $\pm 1/2$, respectively, $p_\mu x^\mu = Et - \boldsymbol{p}\cdot\boldsymbol{x}$, $E = \sqrt{\boldsymbol{p}^2 + m^2}$, $m$ is the mass of the Dirac electron, $u(p,s)$ and $v(p,s)$ are the 4×1 Dirac spinors in the momentum representation. The expansion coefficients $\hat{c}^\dagger$ and $\hat{c}$ (or $\hat{d}^\dagger$ and $\hat{d}$) represent the electron's (or positron's) creation and annihilation operators, respectively. Starting from the Dirac equation $\hat{H}\psi = i\partial\psi/\partial t$ with $\hat{H} = \boldsymbol{\alpha}\cdot\hat{\boldsymbol{p}} + \beta m$ ($\hat{\boldsymbol{p}} = -i\nabla$, $\boldsymbol{\alpha}$ and $\beta$ are the Dirac matrices) one can obtain the equation of continuity, where $\boldsymbol{j} = \psi^\dagger \boldsymbol{\alpha}\psi$ represents the 3D current density of the Dirac field, and then the 3D current operator of the Dirac field is ($\psi^\dagger$ denotes the hermitian conjugate of $\psi$, and so on)

$$\boldsymbol{V} \equiv \int \psi^\dagger \boldsymbol{\alpha}\psi d^3 x = \int \boldsymbol{j} d^3 x. \qquad (2)$$

Let $k^\mu = (\omega, \boldsymbol{k})$ represent the 4D wavenumber vector with $\omega = \sqrt{\boldsymbol{k}^2 + m^2}$ being the corresponding frequency. In the natural units of $\hbar = c = 1$, one has $\boldsymbol{p} = \boldsymbol{k}$ and $E = \omega$. From now on the 4D momentum $p^\mu = (E, \boldsymbol{p})$ in Eq. (1) is rewritten as $k^\mu = (\omega, \boldsymbol{k})$, and then $p_\mu x^\mu = k_\mu x^\mu = \omega t - \boldsymbol{k}\cdot\boldsymbol{x}$. Using Eqs. (1) and (2) one can prove that

$$\boldsymbol{V} = \boldsymbol{V}_{\text{classic}} + \boldsymbol{Z}_\perp + \boldsymbol{Z}_\parallel, \qquad (3)$$

where

$$V_{classic} = \sum_{k,s} (k/\omega)[\hat{c}^\dagger(k,s)\hat{c}(k,s) - \hat{d}^\dagger(k,s)\hat{d}(k,s)] \qquad (4)$$

is the classic current, while $Z_\perp$ and $Z_\parallel$ are the transverse and longitudinal ZB currents, respectively, they are,

$$Z_\perp = \sum_k \{\sqrt{2}\eta(k,1)[\hat{c}^\dagger(k,2)\hat{d}^\dagger(-k,1)\exp(i2\omega t) - \hat{c}(-k,1)\hat{d}(k,2)\exp(-i2\omega t)] + h.c.\}, \qquad (5)$$

$$Z_\parallel = \sum_k (m/\omega)\eta(k,0)\{[\hat{c}^\dagger(k,1)\hat{d}^\dagger(-k,1) - \hat{c}^\dagger(k,2)\hat{d}^\dagger(-k,2)]\exp(i2\omega t) + h.c.\}, \qquad (6)$$

where h.c. denotes the hermitian conjugate of the preceding term, $\hat{c}(\pm k, s) \equiv \hat{c}(k_0, \pm k, s)$ (and so on), and

$$\begin{cases} \eta(k,1) = \eta^*(k,-1) = \dfrac{1}{\sqrt{2}|k|}(\dfrac{k_1 k_3 - ik_2|k|}{k_1 - ik_2}, \dfrac{k_2 k_3 + ik_1|k|}{k_1 - ik_2}, -(k_1 + ik_2)) \\ \eta(k,0) = \dfrac{k}{|k|} = \dfrac{1}{|k|}(k_1, k_2, k_3) \end{cases}, \qquad (7)$$

where $\eta^*(k,-1)$ denotes the complex conjugate of $\eta(k,-1)$ (and so on). In fact, $\eta(k,\pm 1)$ and $\eta(k,0)$ together form a 3D orthonormal basis, they are actually the spinor representation of three linear polarization vectors. Because of $Z_\perp, Z_\parallel \propto \hbar$ (note that $\hbar = c = 1$), $Z_\perp$ and $Z_\parallel$ are the quantum corrections superposing the classic term of $V_{classic}$. Eq. (4) shows that the classical current $V_{classic}$ is formed by electrons or positrons with the momentum $p = k$. In Eqs. (5) and (6), $\hat{c}^\dagger \hat{d}^\dagger$ and $\hat{c}\hat{d}$ are respectively the creation and annihilation operators of electron-positron pairs with vanishing total momentum, then the ZB currents $Z_\perp$ and $Z_\parallel$ are related to the creation and annihilation of virtual electron-positron pairs. In fact, let $A_\mu$ be a 4D electromagnetic potential, $j^\mu = (j^0, j)$ be a 4D current-density vector, according to QED, in the electromagnetic interaction $j^\mu A_\mu$ (let the unit charge $e = 1$), the classical current $V_{classic}$ can contribute to the Compton scattering, while the ZB currents $Z_\perp$ and $Z_\parallel$ can contribute to the Bhabha scattering. However, in the presence of the electromagnetic interaction, the vacuum is replaced with electromagnetic fields, and the electron-positron pairs in the Bhabha scattering are real rather than virtual ones.

Seeing that a hole in Dirac's hole theory can be interpreted as a positron, within the framework of quantum field theory, the traditional argument [12, 13] for the ZB of an electron (in a bound or free state) can be restated as follows: around an original electron, virtual electron-positron pairs are continuously created (and annihilated subsequently) in vacuum, the original electron can annihilate with the positron of a virtual pair, while the electron of the virtual pair which is left over now replaces the original electron, by such an exchange interaction the ZB occurs. Therefore, from the point of view of quantum field theory, the occurrence of the ZB for an electron arises from the influence of virtual electron-positron pairs (or vacuum fluctuations) on the electron. The exchange interaction resulting in the Dirac electron's ZB is shown in Figure 1.

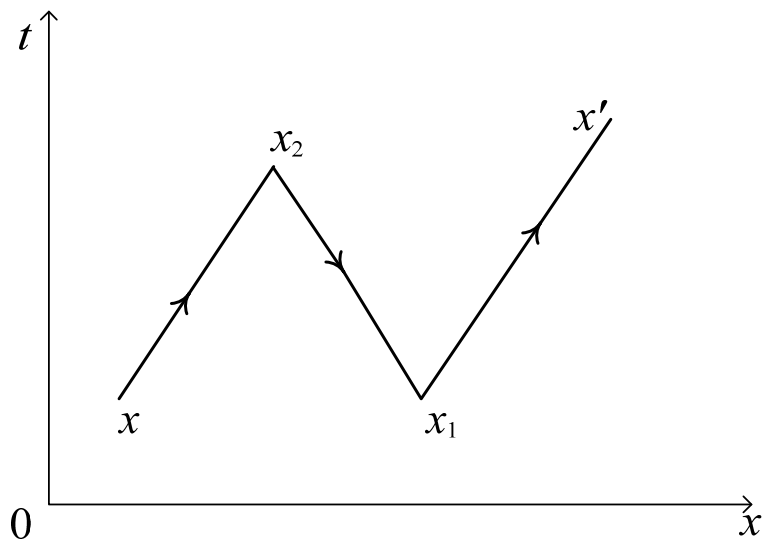

Fig. 1 Exchange interaction resulting in the ZB of an electron

In Figure 1, around an original electron with the initial position of $x$, a virtual electron-positron pair is created at position $x_1$ in the vacuum, its positron annihilates with the original electron at position $x_2$, while its electron arriving at position $x'$ is left over and then replaces the original electron, by such an exchange interaction the ZB of the electron occurs.

## 3. Depicting Hawking radiation in terms of ZB

As for the occurrence of ZB, the effect of electromagnetic vacuum fluctuations can be replaced by that of gravitational vacuum fluctuations. For our purpose, we regard free fields in a curved spacetime as the fields interacting with a gravitational field.

As we know, the Hawking radiation can be viewed from very different perspectives. Now, we present another perspective on the Hawking radiation by means of ZB. Based on Figure 1, let us assume that there is a horizon of a Schwarzschild black hole between positions $x_1$ and $x_2$, and rewrite the $x$-axis as the $r$-axis in spherical coordinates $\{t, r, \theta, \varphi\}$, for the moment Figure 1 becomes Figure 2, where $r_g$ denotes the Schwarzschild radius of the black hole. As we know, quantum theory predicts that black holes emit particles moving away from the horizon. The particles are produced out of vacuum fluctuations of quantum fields present around the black hole [14-16]. As an example, let us assume that a virtual electron-positron pair is created at position $r_1 > r_g$, where the pair's electron moves to $r' \to +\infty$, while its positron moves to $r_2 < r_g$, such that the black hole obtains a negative-energy positron, or equivalently, it loses a positive-energy electron. As result, one can present the black hole evaporation with an equivalent picture as shown in Fig. 2: now that the process that the black hole absorbs a negative-energy antiparticle is equivalent to that the black hole loses a positive-energy particle, let us assume that from the very beginning there is a positive-energy particle (as the original particle) inside the black hole, its original position satisfies $r < r_g$ naturally, afterward at another position, $r_2 < r_g$ say, it annihilates with the negative-energy antiparticle of a virtual particle-antiparticle pair created at position $r_1 > r_g$, which is equivalent to that the original particle is replaced by the virtual pair's particle that moves away from the horizon to $r' \to +\infty$. That is, the black hole's particle at position $r < r_g$ is vaporized in the form of ZB (comparing Fig. 2 with Fig. 1). Therefore, the Hawking radiation can also be viewed from the perspective of ZB. Though Fig. 2 shows that $r < r_2 < r_g$, one may also have $r_2 < r < r_g$.

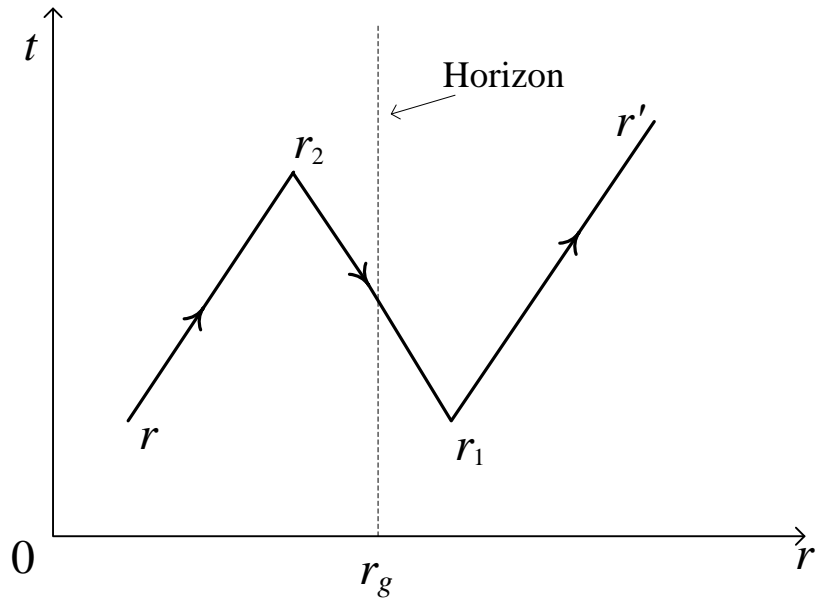

Fig. 2 An equivalent picture of the Hawking radiation

4. **Conclusions**

At the level of quantum field theory, the ZB of the Dirac electron can be studied more exactly, by which one can show that the ZB is due to vacuum fluctuations (e.g., an electromagnetic or gravitational vacuum fluctuations), and its vector property is described by the spinor representation of three linear polarization vectors. In particular, Hawking radiation can also be viewed from the perspective of ZB.

**Acknowledgements**

This work was supported by the Emerging Subject and Innovation Research Fund for the Central Universities (Grant No: ZYGX2010X013).